\documentclass[11pt, a4paper]{article}

\skip\footins=8mm plus 1mm

\usepackage{latexsym,amsmath,amsfonts,amssymb,amsthm}
\usepackage{color}
\usepackage{setspace}
\usepackage{longtable, booktabs}
\usepackage{supertabular}
\usepackage{rotating}
\usepackage{multicol}
\usepackage{multirow}

\usepackage{hyperref}
\usepackage[labelsep=period]{caption}

\newtheorem{theorem}{Theorem}[section]

\newtheorem{remark}[theorem]{Remark}
\newtheorem{definition}[theorem]{Definition}

\def\G1{G^\mathcal{C}}

\long\def\delete#1{}

\numberwithin{equation}{section}
\newcommand{\beq}{\begin{equation}}
\newcommand{\eeq}{\end{equation}}

\newcommand{\be}{\begin{equation}}
\newcommand{\ee}{\end{equation}}
\newcommand{\bea}{\begin{eqnarray}}
\newcommand{\eea}{\end{eqnarray}}
\newcommand{\bean}{\begin{eqnarray*}}
\newcommand{\eean}{\end{eqnarray*}}

\def\G{{\rm G}}

\begin{document}

\title{We can't hear the shape of drum: revisited in 3D case}

\author{\renewcommand{\thefootnote}{\arabic{footnote}} Xiao Hui Liu\footnotemark[1], Jia Chang Sun\footnotemark[1], Jian Wen Cao\footnotemark[1]}

\footnotetext[1]{State Key Laboratory of Computer Science, Institute of Software,
 Chinese Academy of Sciences,  Beijing 100190, P.R. China}

\renewcommand{\thefootnote}{}
\footnotetext{{\em E--mail addresses}: \texttt{xiaohui2014@iscas.ac.cn}, \texttt{jiachang@iscas.ac.cn}, \texttt{jianwen@iscas.ac.cn}}

\date{}

\openup 0.2\jot
\maketitle

\begin{abstract}
``Can one hear the shape of a drum?''was proposed by Kac in 1966. The simple answer is NO as shown through the construction of iso-spectral domains. There already exists 17 families of planar domains which are non-isometric but display the same spectra of frequencies. These frequencies, deduced from the eigenvalues of the Laplacian, are determined by solving the wave equation in a domain, which is subject to Dirichlet boundary conditions. This paper revisits the serials of reflection rule inherent in the 17 families of iso-spectral domains. In accordance with the reflection rule visualized by ``red-blue-black'', we construct real 3D iso-spectral models successfully. What's more, accompanying with the proof of transplantation method,
 we also use the numerical method to verify the iso-spectrality of the 3D models.
\smallskip


\end{abstract}

\section{Introduction}
\label{sec:introduction}
  The famous question as to whether the shape of a drum can be heard has existed for half a century\cite{M.Kac1966USA}, which deduces to the eigenvalues problem of a vibrating membrane in the Euclidean space. Thus if you know the frequencies at which a drum vibrates, can you determine its shape? As we know, the vibration of a drum which spans a domain $\Omega$ in $\mathbb{R}^2$  is governed by the wave equation with Dirichlet boundary
\begin{equation}\label{Equation:wave}
\frac{\partial^2\nu}{\partial t^2}-\left(\frac{\partial^2\nu}{\partial x^2}+\frac{\partial^2\nu}{\partial y^2}\right)=0,~~~~~~~\text{in}~~ \Omega, \\\\
\end{equation}
\begin{equation}\label{Equation:Dboundary}
~~~~\nu=0,~~~~~~~~~~~~~~~~~~~~~~~~~~~~~~~~\text{on}~~ \partial\Omega, \\\\
\end{equation}
where $\nu=\nu(x,y,t)$ denotes the transverse displacement of a point $(x,y)$ in $\Omega$ at time $t$. Seeking a solution by separation of variables $\nu(x,y,t)=F(t)u(x,y)$, we obtain the stationary equation
\begin{equation}\label{Laplace}
\Delta u+\lambda u=0,~~~~~~~~\text{in}~~ \Omega, \\
\end{equation}
\begin{equation}
~~~u=0,~~~~~~~~~~~~~~~~~\text{on}~~ \partial\Omega.
\end{equation}
It is classical that there is an ordered set $\{\lambda_{k},~k=1,2,\ldots\}$ known as eigenvalue spectrum.
\begin{align}
0<\lambda_{1}<\lambda_{2}\leq\ldots\leq\lambda_{k}\ldots;~~\lambda_{k}\rightarrow +\infty,~\text{if}~k\rightarrow +\infty.
\end{align}
The value $\lambda_{k}$ relates to the frequency of the drum with corresponding eigenfunction $u_{k}$. Two bounded domains, $\Omega_{1}$ and $\Omega_{2}$, which have the same set of eigenvalues are called iso-spectral. Additionally, $\Omega_{1}$ and $\Omega_{2}$ are termed isometric if they are congruent in the sense of Euclidean geometry. It is the popularization of Kac's question as to whether there can exist two iso-spectral but non-isometric domains in the real face. It wasn't until 1992 that Gordon et al. \cite{Gordonetal.1992a} answered negatively by finding a pair of non-isometric planar domains with the same Laplace spectrum.

Actually at Kac's time it was known that the answer is NO in the realm of Riemannian manifolds. J. Milnor had constructed two flat tori of dimension 16 which are iso-spectral but not isometric \cite{J.Milnor1964USA}. In the ensuing 25 years many examples of iso-spectral
manifolds mathematically were found, whose dimensions, topology, and curvature properties were discussed. For more detailed specification, please see \cite{Ikeda1980,Vigneras1980,Urakawa1982,Protteret.al1987,Melrose1983,Osgoodetal.1988a,Osgoodetal.1988b}. The crucial step was achieved in 1985 by Sunada \cite{Sunada1985Ann.Math}. The simplest, but abstract theorem gives a sufficient condition for building pairs of iso-spectral manifolds. So, the paradigmatic example given by Gordon et al. was also generated from Sunada's theorem. In fact, the other examples which were constructed after 1992 all followed Sunada's theorem. Buser et al. gave 17 families of iso-spectral pairs in planar case, and a particularly simple method called transplantation technique for detecting iso-spectrality \cite{P.Buser.et.al1994}. Interestingly, Chapman has visualized the transplantation map in terms of ``paper-folding'' \cite{S.J.Chapman1995}. With the paper-folding method, it is clear that what matters is the way how the building blocks are glued together (i.e. reflection rule), irrespective of their shape. More exotic iso-spectral shapes can be made just following the same patten of reflection rule.

It should be remarked that the transplantation method is not independent of the Sunada's theorem. In a summary paper \cite{Brooks}, Brooks present four proofs of the Sunada's theorem including the alternative
proof based on the transplantation method. Continuously, Okada and Shudo employed the transplantation method to verify the equivalence between iso-spectrality and iso-length \cite{Y.Okadaet.al2001}. 17 families of iso-length graph called ``transplantation pairs'' were enumerated to compare with the preexisting planar iso-spectral pairs. Following the edged-graph proposed in \cite{Y.Okadaet.al2001}, theoretical physicist Giraud made a C program to exhaust all possible pairs numerically \cite{O.Giraudet.al2004France}\cite{O.Giraudet.al2005France}. He blazed a trial combining with the finite projective plane approach to explain the reasons for the existence of iso-spectral pairs. Still, essentially only 17 families of examples that say no to Kac's question were constructed in a 50 year period. Thus far, all examples of non-congruent iso-spectral domains in Euclidean space are non-convex. Please see \cite{{Gordonetal.1994a,Gordonetal.1994b}} for the convex domain in hyperbolic plane. Giraud and Thas \cite{O.Giraudet.al2010France} have done good job by reviewing mathematical and physical aspects of iso-spectrality, covering wide range of knowledge  including also pioneering contributions of the authors.

Although iso-spectrality is proved on mathematical grounds, the knowledge of exact eigenvalues and eigenfunctions can not be obtained analytically for such systems. Experiments including numerical investigation and physical implementation contribute a lot.
Early in 1994, Wu and Sprung \cite{Wuet.al1995Canada} verified the iso-spectrality of the first pair (termed as GWW ) numerically by an extrapolated mode-matching method. The ``analytical'' 9$th$ and 21$st$ modes there, corresponding to known simple modes of the underlying triangles, were not only emphasized but were taken at their exact values to 4-5 digits. Subsequently, Driscoll \cite{T.ADriscoll1997}, using a much more accurate modified domain-decomposition method \cite{Desclouxet.al1983}, verified iso-spectrality to 12 significant digits for the first 25 modes, including the two ``analytical'' modes. In 2005, the same or better accuracy was achieved by Betcke and Trefethen with a simple approach \cite{T.Betckeet.al2005}. They modified the famous method--Method of Particular Solutions \cite{L.Foxet.al1967}, and revived it to calculate eigenvalues and eigenmodes of the Laplacian in planar regions.

On the experimental side, Sridhar and Kudrolli \cite{Scridharet.al19931994} performed measurements
on thin microwave cavities shaped like GWW. The measured spectrum has subsequently been confirmed by Wu's numerical simulation \cite{Wuet.al1995Canada}. Then Even and Pieranski \cite{Evenet.al1999} constructed actual shaped small ``drums''--membranes made from liquid crystal smectic films, and measured their vibrations.  To see earlier physical results, please refer \cite{Dharet.al2003,Gottlied2004,Knowleset.al2004} and references therein. More recently, Moon et al. \cite{Moonet.al2008} utilized the iso-spectrality of the electronic nanostructures to extract the quantum phase distributions. Interestingly, they checked that one could indeed ``hear'' iso-spectrality by converting the average measured spectra into audio frequencies. For more details, please see the movie caption in \cite{Moonet.al2008}. Later, phase extraction in disordered iso-spectral shapes was numerically emphasized in \cite{Mugurel2012}. Furthermore, P. Amore \cite{P.Amore2013} extended the iso-spectrality to a wider class of physical problems, including the cases of heterogeneous drums and of quantum billiards in an external field.

\begin{figure}[!htp]
\begin{center}
{\includegraphics[width=1\textwidth]{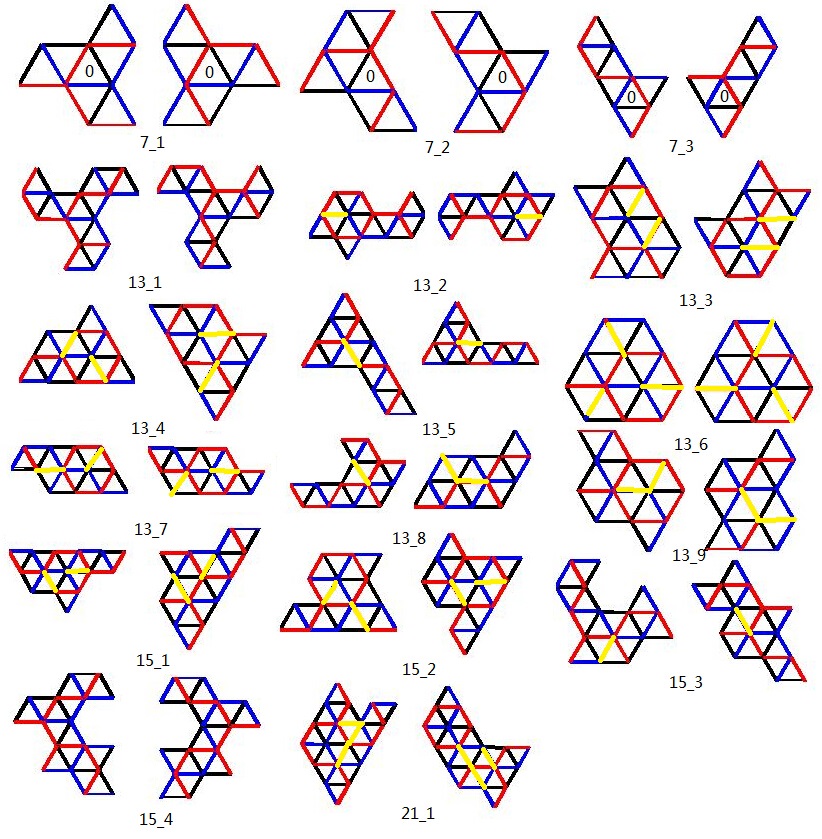}}
\vspace{-0.5cm}
\caption{\small{17 families of iso-spectral pairs constructed by Buser et al. \cite{P.Buser.et.al1994}. The gallery of examples represents not a single pair of iso-spectral domains, but a whole family of pairs of iso-spectral domains, gotten by replacing the equilateral triangles with general triangles. It is not the shape of basic building block, but the series of reflection which is important. The mirror reflection rule is visualized in color red-blue-black. The yellow edge means the overlapped boundary edge. For more special properties (e.g. overlapping, isometric), see the appendix in review \cite{O.Giraudet.al2010France}.}}\label{Fig:17}
\end{center}
\end{figure}

The most basic known planar iso-spectral domains (GWW) are nine-sided polygons that
have also been termed as Bilby and Hawk \cite{Driscoll2003}. Each is a polyform composed
of seven identical triangles and created via a specific series of reflections, such that every triangle is the mirror image of its neighbors. Actually, they are a simplified version of the pair $7_{3}$. Previously based on the reflection rule of $7_{3}$, Cox \cite{Cox2012} constructed three dimensional iso-spectral domains. In addition, 45 shapes assembled by unit cube, square-based prism and right-angled wedge  were depicted to show non-isospectrality in Moorhead's Ph.D. thesis \cite{Moorhead2012}. While the GWW solid iso-spectral prisms (please also see \cite{Mugurel2012}) take the equal eigenvalues, they are just constructed from GWW by sweeping the face into a third orthogonal direction for the same distance. We revisit the serials of reflection rule inherent in the 17 families of iso-spectral domains, and visualize the rule in color red-blue-black. Figure \ref{Fig:17} depicts the specific shape in 2D case.  In accordance with the reflection rule, we successfully construct real 3D iso-spectral models assembled by seven tetrahedrons. The eigenfunctions of our 3D models are not only transplantable, but also numerically equal .

In Section 2 we will study the transplantation method that guarantees the isospectrality of the 17 planar pairs. In Section 3 we show how to construct the iso-spectral models in 3D space. In Section 4 we present some
visualized 3D pairs which are iso-spectral and non-isometric. Finally, conclusion remark briefly shows that the extension of constructing iso-spectral pairs can also be used in higher dimensions.

\newpage
The classifications are illustrated below:
\begin{itemize}
\item 7 tiles with 3 pairs: ~$7_{1}$,  $7_{2}$, $7_{3}$; 
\item 13 tiles with 9 pairs: ~$13_{1}$,  $13_{2}$, $\cdots$, $13_{9}$;
\item 15 tiles with 4 pairs: ~$15_{1}$,  $15_{2}$, $15_{3}$, $15_{4}$;
\item 21 tiles with 1 pairs: ~$21_{1}$.
\end{itemize}
\newpage
\section{Transplantation method}\label{Sec:Transplantation method}
\noindent
The transplantation proof was first applied to Riemann surfaces by
Buser \cite{P.Buser1986Switzerland}, and to
 reexamine iso-spectrality of planar domains given in \cite{P.Buser.et.al1994}. Quoting sentence from Buser ``We shall show that the eigenfunctions on the first surface can be suitable transplanted to yield eigenfunctions with the same eigenvalue on the second surface and vice-versa'', we know suitable transplantation is important.
 B\'{e}rard \cite{P.Berard19921993} followed this versatile method to reproof the iso-spectrality of the pair constructed by Gordon et al.. Okada and Shudo \cite{Y.Okadaet.al2001} formalised the matrix representation (i.e. transplantation matrix $T$) of transplantation of eigenfunctions. In fact, it turns out that the matrix $T$ is just the incidence matrix of the graph
associated with a certain finite projective space \cite{O.Giraudet.al2004France} \cite{O.Giraudet.al2005France}. Giraud and Thas \cite{O.Giraudet.al2010France} reviewed the fascinating relation between the iso-spectral pairs and geometry of vector spaces over finite fields. Illustrating one simple method to obtain transplantation matrix of the 17 pairs, we follow the definition mentioned in \cite{Y.Okadaet.al2001,O.Giraudet.al2004France,O.Giraudet.al2005France,O.Giraudet.al2010France}, and describe it with red-blue-black reflection rule.
\begin{figure}[!htp]
\begin{center}
{\includegraphics[width=1\textwidth]{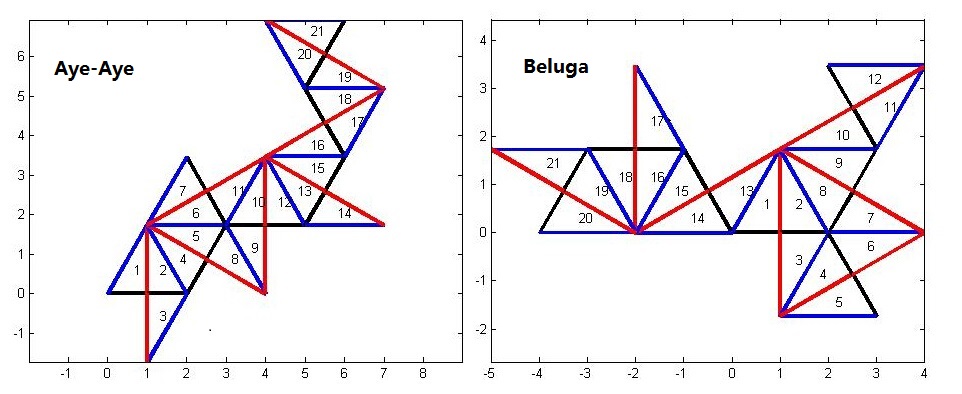}}
\vspace{-0.5cm}
\caption{\small{The special pair of $21_{1}$: termed as Aye-Aye and Beluga \cite{Moonet.al2008}. Each is a polyform composed of 21 identical triangles with angles $(30^o,60^o,90^o)$, and created via a specific series of mirror reflections (e.g. basic tile labeled 1 is mirrored to tile 2 by red side). The labels can be arbitrary, for convenience, we keep them consistent with the labels in \cite{Moonet.al2008}.}  }\label{Fig:21}
\end{center}\vspace{-0.2cm}
\end{figure}
\begin{definition}\label{Def:}
\rm The transplantation matrix $T$ must be such that
 \begin{equation}\label{Equation:TABT}
T A^{\nu} = B^{\nu} T,~\nu=1~(\text{red}),~~2~(\text{blue}),~3~(\text{black}).
\end{equation}
In Equation (\ref{Equation:TABT}), matrices $A^{\nu}$ and $B^{\nu}$ describe how the tiles are glued togethers (i.e. reflection rule). For example,
\begin{itemize}
\item $A^{\nu}_{i,j}=1$ if and only if the edge number (color) $\nu$ of $i$ glues tile $i$ to tile $j$;
\item $A^{\nu}_{i,i}=-1$ if the edge number $\nu$ of $i$ glues nothing (boundary);
\item 0, otherwise.
\end{itemize}
\end{definition}

For special pair of $21_{1}$ depicted in Figure \ref{Fig:21} , $\nu$ runs form 1 to 3 (red to black). That is also alternatively corresponding with middle, long and short edge. If you keep the labels of Aye-Aye and Beluga shape, line $i$ contains a 1 at the column $j$ where $i$ and $j$ are glued by their red side; Otherwise the red side is the boundary of the tile, line $i$ contains a -1. So for the red side, the matrix $A^1$ (associated with the Aye-Aye shape) would be that
$A^1_{1,2}=1,~A^1_{2,1}=1,~A^1_{3,3}=-1,~A^1_{4,5}=1,~A^1_{5,4}=1$ and so on.
Similarly the $B^1$ matrix would be associate with the Beluga shape and yields $B^1_{1,2}=1, B^1_{2,1}=1,\ldots,B^1_{6,6}=-1$ and so on. Then once you have all matrices $A^{\nu}$ and $B^{\nu}$ for the three sides, finding $T$ just amounts to solving a system of equation.
\begin{figure}[!htp]
\begin{center}
{\includegraphics[width=1\textwidth]{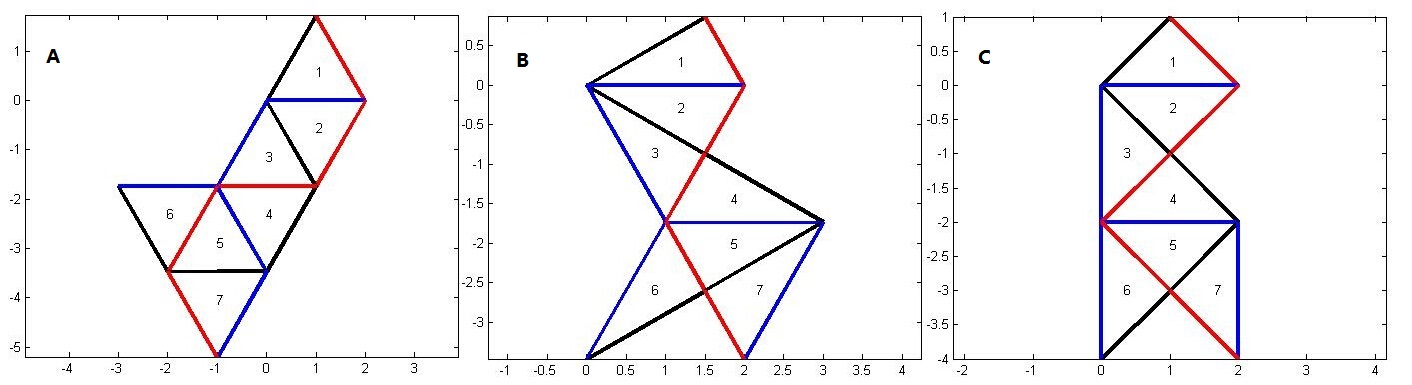}}
\hfill
\includegraphics[width=1\textwidth]{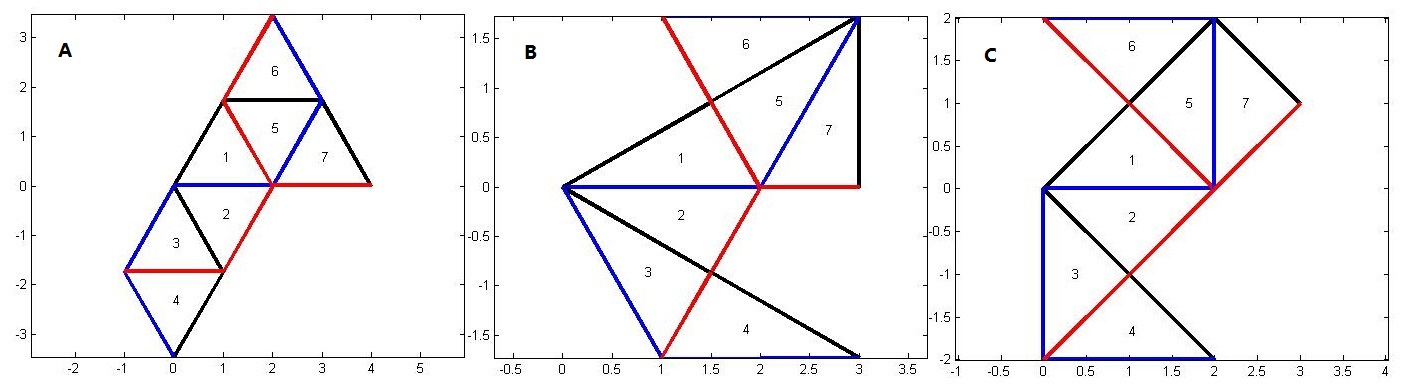}
\caption{\small{Three iso-spectral pairs following the reflection rule of $7_{3}$ (e.g. basic tile labeled 1 is mirrored to tile 2 by blue side). (\textbf{A}) This pair is isometric with equilateral triangle as the basic tile. (\textbf{B}) This pair is non-isometric with $(30^o,60^o,90^o)$ triangle as the basic tile. Sleeman and Hua \cite{SleemanChen2000} utilized it to construct iso-spectral domains with fractal border. (\textbf{C}) This  pair (GWW) is the paradigmatic pair constructed by Gordon et al.}  }\label{Fig:73}
\end{center}
\end{figure}
If we keep the notations that $T:=\{C_{ij}\}_{N\times N}$ ($N$ represents 7, 13, 15, 21), combining reflection rule in Figure \ref{Fig:73} with definition \ref{Def:}, $T$ can be calculated in the following form:
\begin{equation}\label{Equation:T}
 T=\left(
      \begin {array} {ccccccc}
       C_{11} & -C_{13} & C_{13} & - C_{13} & -C_{11} & C_{11} & C_{13} \\
       - C_{13} & C_{11} & -C_{13} & C_{13} & C_{13} & - C_{11} & - C_{11} \\
        C_{13} & - C_{13} & C_{11} & - C_{13} & - C_{11} & C_{13} & C_{11} \\
        - C_{11} & C_{13} & - C_{13} & C_{11} & C_{13} & - C_{13} & - C_{11} \\
         C_{13} & - C_{11} & C_{13} & - C_{11} & - C_{11} & C_{13} & C_{13} \\
          - C_{11} & C_{11} & - C_{11} & C_{13} & C_{13} & - C_{13} & - C_{13} \\
          - C_{13} & C_{13} & - C_{11} & C_{11} & C_{13} & - C_{11} & - C_{13} \\
            \end {array}
              \right).
\end{equation}
$C_{11}$ and~$C_{13}$ are the two remaining degrees of freedom after solving Equations (\ref{Equation:TABT}). Extracting $C_{11}$ and~$C_{13}$ from Equation (\ref{Equation:T}), hence $T$ is simplified like this
\begin{equation}\label{Equation:T3T4}
T=C_{11}T_{3}+C_{13}T_{4},~C_{11}\neq C_{13}.
\end{equation}
$T_{k}$ has only $k$ nonzero entries in each row and each column. In Buser's words \cite{P.Buser.et.al1994}, $T_{3}$ is the original mapping while $T_{4}$ is the complementary mapping. Any linear combination like Equation (\ref{Equation:T3T4})
will also be a transplantation mapping. Once $C_{11}$ is equal to $C_{13}$, there will be the nodal line case (i.e. eigenfunctions will vanish, see book \cite{RCourantHilbert} for the explaination), in which the ``analytical'' solutions appear.

One can show that transplantation is a sufficient condition to guarantee iso-spectrality (if the matrix $T$ is not merely a permutation matrix, in which case the two domains would just have the same shape). The underlying idea is that if $\phi$ is an eigenfunction of the first domain and $\phi_{i}$ its restriction to tile $i$, then one can build an eigenfunction $\varphi$ of the second domain as
\begin{equation}\label{eigenfunction}
\varphi_{i}=\mathcal{C}\sum_{j=1}^{N}T_{ij}\phi_{j},~i=1,2,\ldots,N.
\end{equation}
$\mathcal{C}$ is some normalization factor in Equation (\ref{eigenfunction}). The smoothness of $\varphi_{i}$ can be easily checked on all reflecting sides and zero boundary conditions of the second domain, if the above conditions are satisfied.
Following this procedure, the two domains discussed will then be iso-spectral, since the same map transposing eigenfunctions of the second domain onto the first domain will work as before.  For more mathematical proof, please see pp. 8 in review \cite{O.Giraudet.al2010France}. Simultaneously, you can also refer Chapman's paper-folding for the visualization of transplantation.

As previously described, it is not even necessary for the basic tile to be a triangle. Arbitrary shape with at least three edges will work. We simply choose a triangle with three sides to represent the three edges, on which we will reflect the shape. Experiments drawn on frames of arbitrary shapes were conducted by Even and Pieranski \cite{Evenet.al1999} using vibrations of smectic films. Meanwhile, two series of reflections generating from $7_{3}$ were highly emphasized. Taking the words of Even and Pieranski, If one broke the reflection rule, the consequences on the spectrum are drastic. Moon et al. \cite{O.Giraudet.al2010France} also designed one ``Broken Hawk'' violating the rule, which led significantly difference compared with the spectrum of Bilby and Hawk. In summary, specific reflection rules inherent in the 17 iso-spectral pairs are very important, when associating with transplantation matrix to transplant eigenfunctions. Although this theorem considers the 2D case only, the 3D
case is similar.

With notations $\alpha$ and $\beta$ meaning formal parameters, the calculating results of transplantation matrix for 17 iso-spectral pairs are given below:
\begin{itemize}
\item 7~tiles:  $T=\alpha T_{3}+\beta T_{4},~\alpha \neq \beta$;
\item 13~tiles: $T=\alpha T_{4}+\beta T_{9},~\alpha \neq \beta$;
\item 15~tiles: $ T=\alpha T_{7}+\beta T_{8},~\alpha \neq \beta$;
\item 21~tiles:  $T=\alpha T_{5}+\beta T_{16},~\alpha \neq \beta$.
\end{itemize}
\begin{remark}\rm
It is easy to obtain similar relations for
Neumann boundary conditions by conjugating all matrices $A^\nu$ and $B^\nu$ in definition \ref{Def:}, with a diagonal
entry $A^\nu_{i,i}=1$ according to whether tile $i$ glues nothing (boundary).
\end{remark}
\section{How to construct the 3D iso-spectral models }
\noindent
In this section, we explain the idea of constructing iso-spectral pairs in 3D space, which also can be seen as a real extension of the 2D pairs. There are a number of options for selecting basic tile in 3D space, such as parallelepiped, tetrahedron, etc. Once you have selected the basic tile (tetrahedron for example), you only need to fix one of its faces (meaning that you just forget about it) and color the three other faces by red, blue and black. Then you do the mirror reflection with respect to red, blue and black according to the rule. That will automatically generate many 3D iso-spectral pairs.
\subsection{\rm Extensions of $7_{1}$ for instance}
\noindent

\begin{figure}[!htp]
\begin{center}
{\includegraphics[width=1\textwidth]{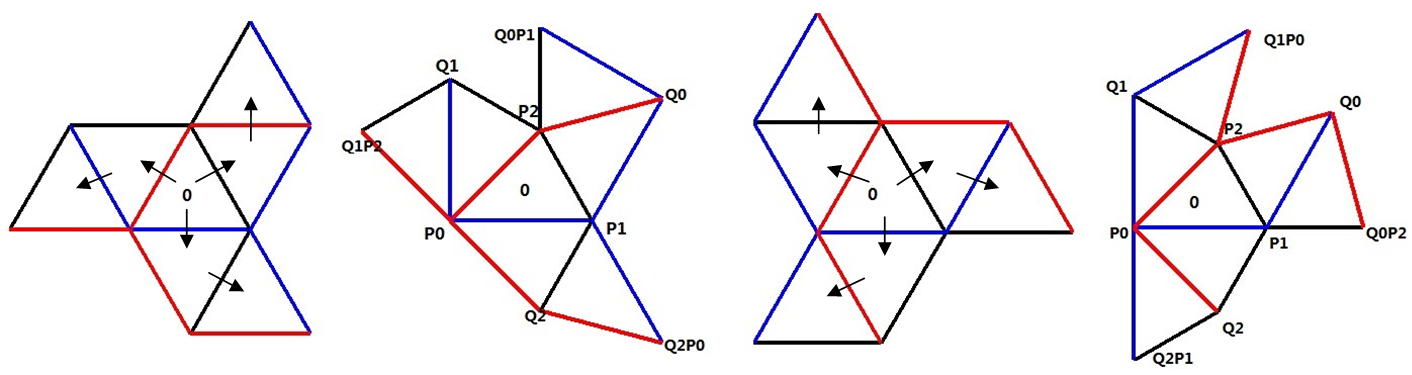}}
\vspace{-0.7cm}
\caption{\small{The propeller and warped propeller iso-spectral pairs. The latter is gotten by replacing the equilateral triangles
with $(45^o,60^o,75^o)$ triangles so that the triangles labelled 0 are mapped onto one
another by a translation. The remaining triangles are obtained from these
by the appropriate sequence of reflections, with the indication of the arrows. For convenience, each vertex of the triangle is labelled, such as $Q_{0}$, $Q_{1}$, etc.} }\label{Fig:propeller}
\end{center}
\end{figure}

In order to facilitate the description of this extension, we first discuss the propeller-shaped example \cite{P.Buser.et.al1994}, generated from the $7_{1}$ pair. In both propellers the central triangle labelled ``0'' has a distinguishing property: its sides connect the three inward corners of the propeller. Based on this property, we choose one basic tetrahedron placing the position ``0'' of Figure \ref{Fig:propeller}. Following the same orientation, we only need to perform the series of reflections rule depicted in Figure \ref{Fig:propeller}. It is remarked that the biggest difference between 2D and 3D is that 2D is reflected by edges, while 3D is reflected by faces. The following steps describe how to build up 3D iso-spectral models.
\begin{itemize}
\item Step 1: Given four 3D vertices to form a basic tetrahedron tile $\{P_{0},P_{1},P_{2},P_{3}\}$.\\
 $P_0=(x_0,y_0,z_0),~P_1=(x_1,y_1,z_1),~P_2=(x_2,y_2,z_2),~P_3=(x_3,y_3,z_3)$.
\item Step 2: Given 3D face $( P_1,P_2,P_3)$, for a point $P_0$, its mirror reflection
mapping can be expressed as $Q_0=\emph{\textbf{Mirror}}(P_0;~P_1, P_2, P_3)$. The relevant calculation formulas are \begin{align}
    Q_0= P_0 + 2 Q_0 P ,~~ Q_0 P= \alpha_1 P_1 +\alpha_2 P_2 +\alpha_3 P_3,
\end{align}\label{Equation:mirror1}
 where  $Q_0 P$ is the projection point of $P_0$ on the face with three constrains 
\begin{align}\label{Equation:mirror2}
   ( Q_0 P-P_0) \cdot  (P_2-P_1)=0,\quad ( Q_0 P-P_0) \cdot  (P_2-P_3)=0,\quad \alpha_1 +\alpha_2 +\alpha_3=1.
\end{align}
\item Step 3: Fix or forget one 3D face (e.g. $(P_0,P_1,P_2)$). Then we can obtain the other two reflection points $Q_{1}$ and $Q_{2}$ with respect to faces $(P_0,P_2,P_3)$ and $(P_0,P_1,P_3)$.
\begin{align}
Q_1 = \emph{\textbf{Mirror}}(P_1;~P_0,P_2,P_3),\quad Q_2 = \emph{\textbf{Mirror}}(P_2;~P_0,P_1,P_3). \notag
\end{align}
\end{itemize}

So far, four tetrahedrons assembled in our 3D model are as follows.
\begin{align}
 \{P_{0},P_{1},P_{2},P_{3}\},~~~\{Q_{0},P_{1},P_{2},P_{3}\},~~~\{P_{0},Q_{1},P_{2},P_{3}\},~~\{P_{0},P_{1},Q_{2},P_{3}\}. \notag
\end{align}
They are all glued by ``colored'' faces. With further multilevel mirror operation, the next step is to construct two models by adding three tetrahedrons in two different ways (Class $7_{1}$ for example).
\begin{itemize}
\item Class $7_{1}$ \#\textbf{A}:
$\{Q_0, Q_0P_1, P_2, P_3\},~~\{P_{0},Q_{1},Q_{1}P_{2},P_{3}\},~~\{Q_{2}P_{0},P_{1},Q_{2},P_{3}\}$;
\small{\begin{align}
Q_0P_1&=\emph{\textbf{Mirror}}(P_1;~Q_0,P_2,P_3),~Q_{1}P_{2}=\emph{\textbf{Mirror}}(P_2;~P_0,Q_1,P_3), ~Q_{2}P_{0}=\emph{\textbf{Mirror}}(P_0;~P_{1},Q_{2},P_{3}).\notag
\end{align}} \normalsize
\item Class $7_{1}$ \#\textbf{B}:
$\{Q_0, P_1, Q_0P_2,P_3\},~~\{Q_{1}P_{0},Q_{1},P_{2},P_{3}\},~~\{P_{0},Q_{2}P_{1},Q_{2},P_{3}\}$;
\small{\begin{align}
Q_0P_2&=\emph{\textbf{Mirror}}(P_2;~Q_0,P_1,P_3),~Q_{1}P_{0}=\emph{\textbf{Mirror}}(P_0;~Q_{1},P_{2},P_{3}), ~Q_{2}P_{1}=\emph{\textbf{Mirror}}(P_1;~P_{0},Q_{2},P_{3}).\notag
\end{align}}
 \end{itemize}\normalsize
\begin{remark}\rm Sometimes, there will be isometric models in 3D space. In Appendix, we present 3D models constructed by seven unit cubes. Because of the symmetry in unit cube, the models in Figure \ref{Fig:cube2} and \ref{Fig:cube3} are isometric. Okada and Shudo \cite{Y.Okadaet.al2001} demonstrated that in 2D case, if one destructs the symmetry by changing the shape of the building block, such an ``accidental'' situation can be avoided. That is also useful in 3D case.
\end{remark}\label{Remark:3.1}
\subsection{\rm Extensions of $7_{2}$, $7_{3}$ for instance}
\begin{figure}[!htp]
\begin{center}
{\includegraphics[width=1\textwidth]{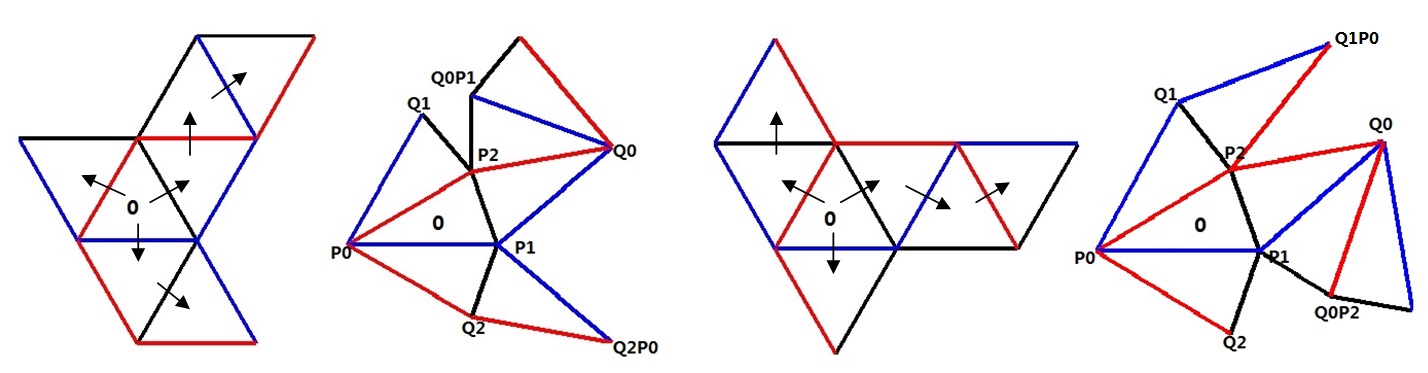}}
\vspace{-0.7cm}
\caption{\small{Two planar $7_{2}$ iso-spectral pairs. The latter is gotten by replacing the equilateral triangles
with $(30^o,70^o,80^o)$ triangles (once appeared in \cite{T.ADriscoll1997}) so that the triangles labelled 0 are mapped onto one
another by a translation. For convenience, each vertex of the triangle is labelled, such as $Q_{0}$, $Q_{1}$, etc.} }\label{Fig:72eg1}
\end{center}
\end{figure}

\begin{figure}[!htp]
\begin{center}
{\includegraphics[width=1\textwidth]{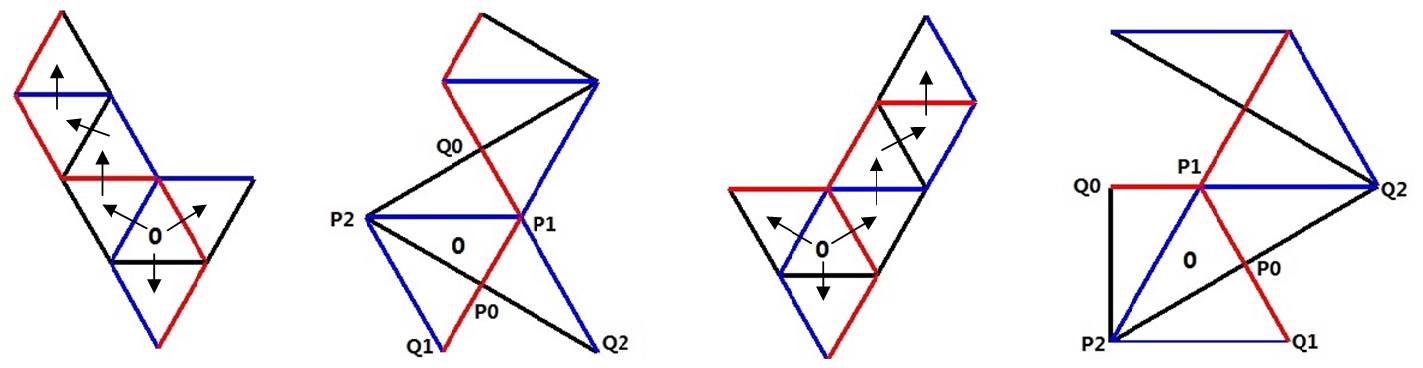}}
\vspace{-0.7cm}
\caption{\small{Two planar $7_{3}$ iso-spectral pairs. The latter is gotten by replacing the equilateral triangles
with $(30^o,60^o,90^o)$ triangles (once appeared in \cite{SleemanChen2000,Moonet.al2008}) so that the triangles labelled 0 are mapped onto one
another by a translation. For convenience, each vertex of the triangle is labelled, such as $Q_{0}$, $Q_{1}$, etc.} }\label{Fig:73eg1}
\end{center}
\end{figure}

Here, following the arrows' indications depicted in Figure \ref{Fig:72eg1}, we can easily construct 3D iso-spectral models as an extension of $7_{2}$. We choose the basic tetrahedron to replae the position ``0'' of Figure \ref{Fig:72eg1}, and perform the reflection rule continuously. Figure \ref{Fig:73eg1} depicts the same situation of $7_{3}$. The remaining tetrahedrons will be obtained with further multilevel mirror operation. For the other 14 families of iso-spectral pairs, their extensions of constructing 3D models are similar with $7_{1}$, $7_{2}$ and $7_{3}$, so we omit it.

\section{3D examples---Basic Simplex Tetrahedron---Numerical tests} 
\noindent In this section, we shall present 3D iso-spectral models constructed by seven Basic Simplex Tetrahedrons ($0\le y\le x\le z\le 1$). Numerical method finite difference is emphasized to verify the iso-spectrality of the 3D models.
 \begin{itemize}
\item Step 1. Given four 3D vertices of the Basic Simplex with coordinate,\\
$P_0= \{0, 0, 0\},\quad  P_1= \{0, 0, 1\}, \quad P_2 = \{1, 0, 1\},\quad  P_3=\{1, 1, 1\};$
\item Step 2. Find three mirror points, \\
$Q_0= \{0, 0, 2\},\quad Q_1= \{1,0, 0\},\quad Q_2 = \{0,1, 1\}$;
\item Step 3. Illustrate the four tetrahedrons assembled,
\small{\begin{itemize}
  \item Simplex 0: $\{P_0,P_1,P_2,P_3\}$,\qquad $0\le y \le x\le z \le 1$;
  \item Simplex 1: $\{Q_0,P_1,P_2,P_3\}$,\qquad $0\le y \le x\le 2-z \le 1$;
  \item Simplex 2: $\{P_0,Q_1,P_2,P_3\}$,\qquad $0\le y \le z\le x \le 1  $;
  \item Simplex 3: $\{P_0,P_1,Q_2,P_3\}$,\qquad $ 0\le x \le y\le z \le 1 $.
 \end{itemize}}\normalsize
\end{itemize}
\begin{figure}[!htb]
\includegraphics[width=0.27\textwidth]{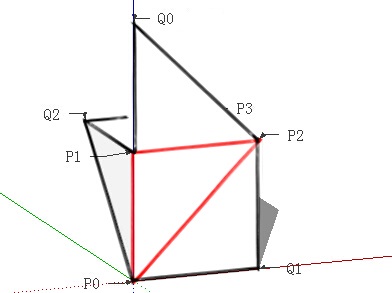}\hspace{-0.7cm}
\includegraphics[width=0.27\textwidth]{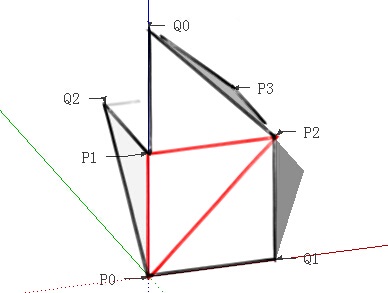}\hspace{-0.5cm}
\includegraphics[width=0.27\textwidth]{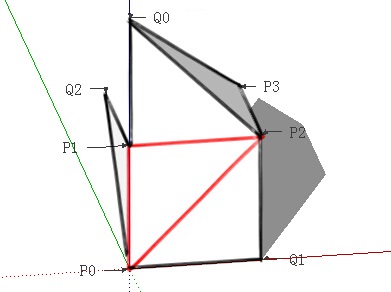}\hspace{-0.5cm}
\includegraphics[width=0.27\textwidth]{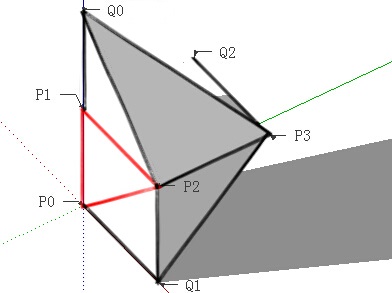}
\hspace*{\fill}\vspace{-0.1cm}
\caption{\small{Four Simplex Tetrahedrons assembled (termed as kernel Simplex), viewing from different angles. Face $(P_0,P_1,P_2)$ is special colored to help us forget it when performing reflection rule. Consequently, faces $(Q_0,P_1,P_2)$, $(P_0,Q_1,P_2)$ and $(P_0,P_1,Q_2)$ are not considered as reflection faces, either.}}\label{kernal simplex}
\end{figure}

\normalsize
 As a foundation of constructing non-isometric 3D models, kernel Simplex (depicted in Fig. \ref{kernal simplex}) plays quite an important role, especially for the extension of $7_{1}$, $7_{2}$ and $7_{3}$. Following the reflection rule, we just conduct the multilevel mirror operation on the kernel Simplex, which is just adding some tetrahedrons. The following step is to classify two non-isometric models generated from $7_{1}$.
\noindent
\begin{itemize}
\item Class $7_{1}$ \#\textbf{A}, three mirror points:
$Q_0P_1=\{1, 0, 2\},~Q_{1}P_{2}=\{1, 1, 0\},~Q_{2}P_{0}=\{0, 0, 2\}.$
\small{\begin{itemize}
\item Simplex 4: $\{Q_0, Q_0P_1, P_2, P_3\}$,\qquad $0\le y \le 2-z\le x \le 1  $;
  \item Simplex 5: $\{P_{0},Q_{1},Q_{1}P_{2},P_{3}\}$,\qquad $0\le z \le y\le x \le 1 $;
  \item Simplex 6: $\{Q_{2}P_{0},P_{1},Q_{2},P_{3}\}$,\qquad $0\le x \le y\le 2-z \le 1  $.
 \end{itemize}}\normalsize
\item Class $7_{1}$ \#\textbf{B}, three mirror points:
$Q_0P_2=\{0, 1, 1\},~Q_{1}P_{0}=\{2, 0, 0\},~Q_{2}P_{1}=\{0, 1, 0\}$.
\small{\begin{itemize}
\item Simplex 4: $\{Q_0, P_1, Q_0P_2,P_3\}$,\qquad $0\le x \le y\le 2-z \le 1  $;
  \item Simplex 5: $\{Q_{1}P_{0},Q_{1},P_{2},P_{3}\}$,\qquad $0\le y \le z\le 2-x \le 1 $;
  \item Simplex 6: $\{P_{0},Q_{2}P_{1},Q_{2},P_{3}\}$,\qquad $0\le x\le z\le y \le 1  $.
 \end{itemize}}\normalsize
 \end{itemize}
 \begin{figure}[!htp]
\includegraphics[width=0.25\textwidth]{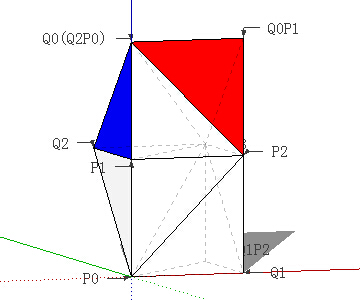}
\includegraphics[width=0.25\textwidth]{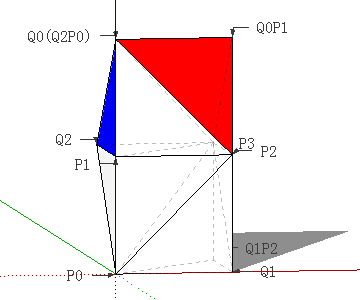}
\includegraphics[width=0.25\textwidth]{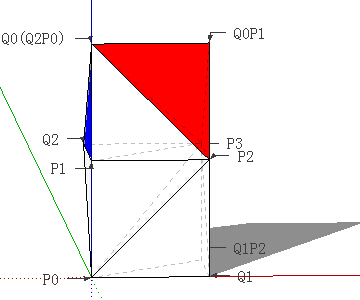}
\includegraphics[width=0.25\textwidth]{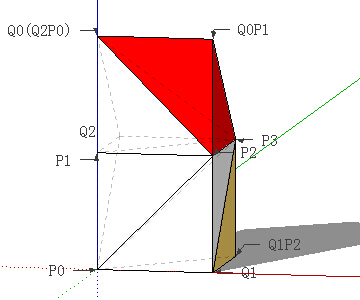}\\
\includegraphics[width=0.25\textwidth]{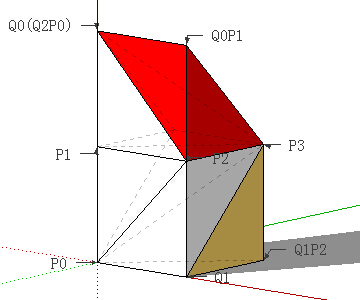}
\includegraphics[width=0.25\textwidth]{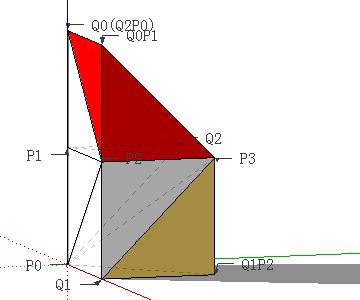}
\includegraphics[width=0.25\textwidth]{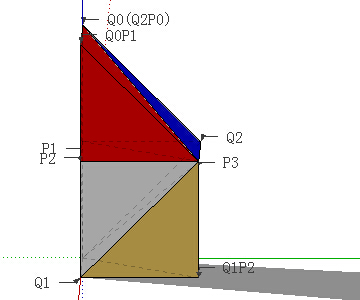}
\includegraphics[width=0.25\textwidth]{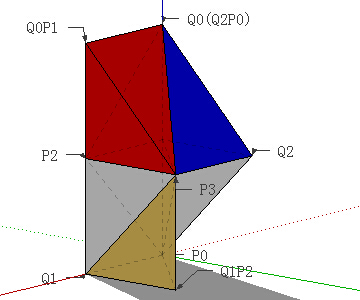}
\hspace*{\fill}\vspace{-0.2cm}
\caption{\small{Seven Simplex Tetrahedrons assembled (termed as Class $7_{1}$ \#\textbf{A}), viewing from different angles. The labels in each vertex help us to distinguish the tetrahedrons. The colored tetrahedrons means the three added tetrahedrons on the Kernel Simplex. Because of the symmetry in Simplex Tetrahedron, points $Q_{0}$ and $Q_{2}P_{0}$ coincide. }}\label{Fig:classA}
\end{figure}
\begin{figure}[!htp]
\includegraphics[width=0.25\textwidth]{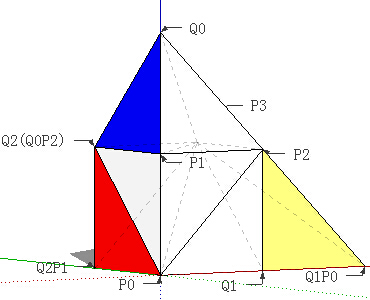}
\includegraphics[width=0.25\textwidth]{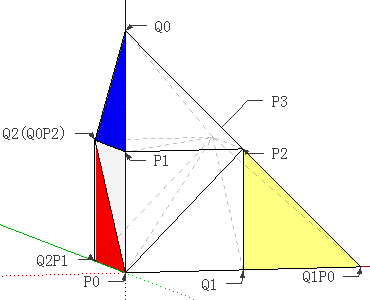}
\includegraphics[width=0.25\textwidth]{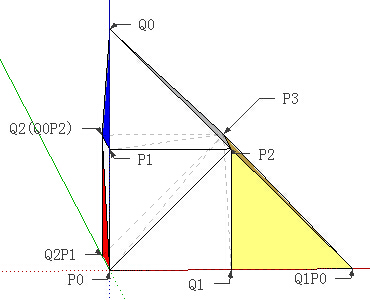}
\includegraphics[width=0.25\textwidth]{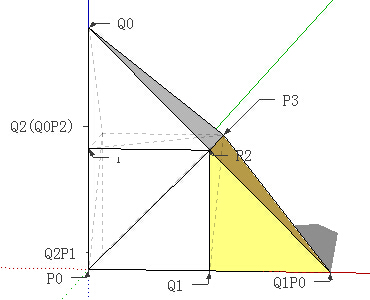}\\
\includegraphics[width=0.25\textwidth]{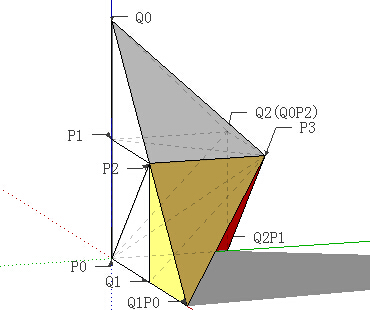}
\includegraphics[width=0.25\textwidth]{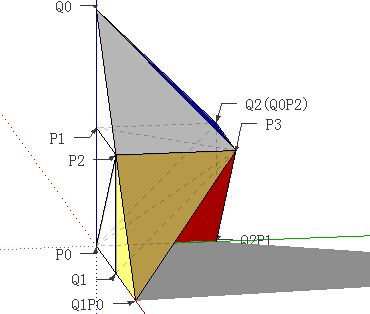}
\includegraphics[width=0.25\textwidth]{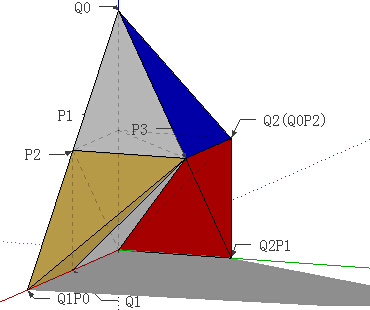}
\includegraphics[width=0.25\textwidth]{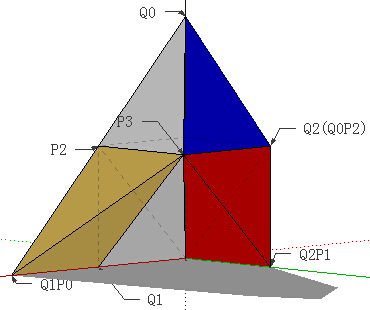}
\hspace*{\fill}\vspace{-0.2cm}
\caption{\small{Seven Simplex Tetrahedrons assembled (termed as Class $7_{1}$ \#\textbf{B}), viewing from different angles. The labels in each vertex help us to distinguish the tetrahedrons. The colored tetrahedrons means the three added tetrahedrons on the Kernel Simplex. Because of the symmetry in Simplex Tetrahedron, points $Q_{2}$ and $Q_{0}P_{2}$ coincide.  }}\label{Fig:classB}
\end{figure}
\vspace{-0.2cm}
\begin{remark}\rm
Figure \ref{Fig:classA} and \ref{Fig:classB} depict one 3D iso-spectral pair termed as Class $7_{1}$ \#\textbf{A} and Class $7_{1}$ \#\textbf{B} respectively. They are all assembled by seven Simplex Tetrahedrons in different ways. Because of the symmetry in the Simplex Tetrahedron, faces $(P_{1},P_{3},Q_{0})$ and $(P_{1},P_{3},Q_{2}P_{0})$ coincide in Class $7_{1}$ \#\textbf{A}. Similarly, faces $(P_{0},P_{3},Q_{2})$ and $(P_{0},P_{3},Q_{0}P_{2})$ coincide in Class $7_{1}$ \#\textbf{B}. For this case, we should also consider these faces as the boundary face.
\end{remark}
\begin{remark}\rm
Class $7_{1}$ \#\textbf{A} and Class $7_{1}$ \#\textbf{B} are non-isometric. With the aid of the 3D Printing technique\footnote[1]{Please see Wiki: https://en.wikipedia.org/wiki/3D\_printing}, we obtain two models which are obviously non-congruent. Moreover, utilizing their distance matrix (the matrix of distances between all vertices), we can also judge they are non-isometric.
\end{remark}
\renewcommand\arraystretch{0.8}
\begin{table}[!htp]
  \small{\begin{center}
  \caption{\small{The first 25 approximated eigenvalues of the 3D iso-spectral pair: Class $7_{1}$ \#\textbf{A} and Class $7_{1}$ \#\textbf{B}, using the simple finite difference method ( please also see \cite{Moorhead2012}) with mesh size $h=1/20$. The Difference refers to the absolute difference between the eigenvalues.}}\vspace{-0.3cm}
    \begin{tabular*}{\linewidth}{@{\extracolsep{\fill}}*{4}{c}}                                   \toprule \toprule
   Eigenvalue    &Class $7_{1}$ \#\textbf{A}   &Class $7_{1}$ \#\textbf{B}   &Difference ($\times 10^{-12}$)   \\\midrule
  1  &  44.4718 &  44.4718 &  0.0568\\
 2  &  62.8210 &  62.8210 &  0.0426\\
 3  &  68.9764 &  68.9764 &       0\\
 4  &  80.4222 &  80.4222 &  0.1279\\
 5  &  86.0231 &  86.0231 &  0.0284\\
 6  & 103.2302 & 103.2302 &  0.1279\\
 7  & 105.6904 & 105.6904 &  0.1847\\
 8  & 110.1293 & 110.1293 &  0.0284\\
 9  & 117.5639 & 117.5639 &  0.3411\\
 10 & 126.6846 & 126.6846 &  0.0853\\
 11 & 130.2792 & 130.2792 &  0.3126\\
 \large{\color{red}12} & 136.1989 & 136.1989 &  0.2842\\
 13 & 136.5769 & 136.5769 &  0.3695\\
 14 & 142.5582 & 142.5582 &  0.6253\\
 15 & 147.9829 & 147.9829 &       0\\
 16 & 154.1811 & 154.1811 &  0.0853\\
 17 & 161.1378 & 161.1378 &  0.1990\\
 18 & 164.5301 & 164.5301 &  0.3979\\
 19 & 169.0497 & 169.0497 &  0.5969\\
 20 & 172.1153 & 172.1153 &  0.3979\\
 21 & 176.0983 & 176.0983 &  0.1705\\
 22 & 180.6497 & 180.6497 &  0.1705\\
 23 & 185.0137 & 185.0137 &  0.0568\\
 24 & 190.4692 & 190.4692 &  0.0284\\
 25 & 194.8700 & 194.8700 &  0.5116\\
  \bottomrule
 \end{tabular*}\label{Table:71}
  \end{center}}
\end{table}

In Moorhead's Ph.D thesis \cite{Moorhead2012}, he present 45 shapes to discuss iso-spectrality, using a simple finite difference scheme to calculate eigenvalues. We are also interested in applying finite difference method to our 3D iso-spectral models. For more detailed results, please see Table \ref{Table:71}. That is if we let $\lambda_{k}$ be the approximated $k$-th eigenvalue for the first Class $7_{1}$ \#\textbf{A}
and $\mu_{k}$ be the approximated $k$-th eigenvalue for the second Class $7_{1}$ \#\textbf{B} then
\[\max_{1\leq k \leq 25}|\lambda_{k}-\mu_{k}|=6.2528 \times 10^{-13}, ~~~\|\lambda_{k}-\mu_{k}\|_{L_{2}}=1.4503 \times 10^{-12}.\]

\begin{figure}[!htp]
\includegraphics[width=0.25\textwidth]{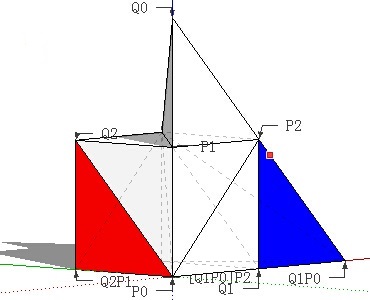}
\includegraphics[width=0.25\textwidth]{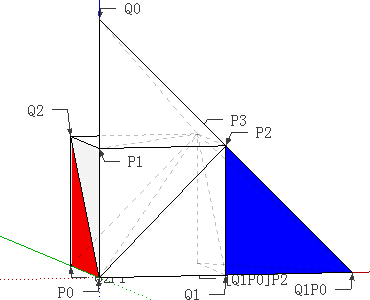}
\includegraphics[width=0.25\textwidth]{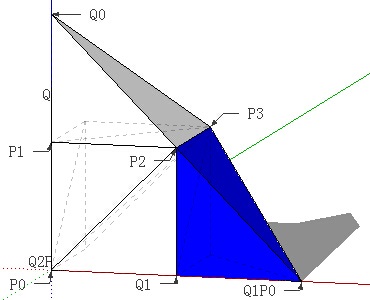}
\includegraphics[width=0.25\textwidth]{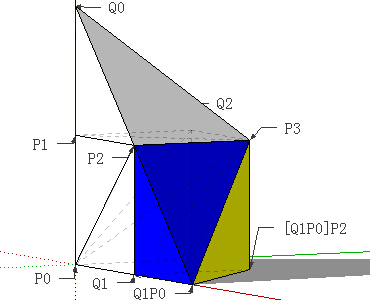}\\
\includegraphics[width=0.25\textwidth]{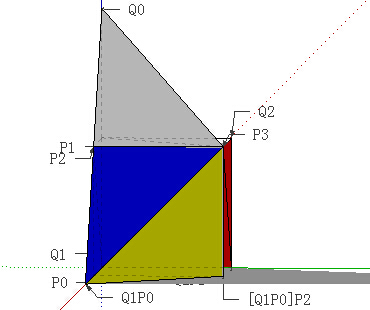}
\includegraphics[width=0.25\textwidth]{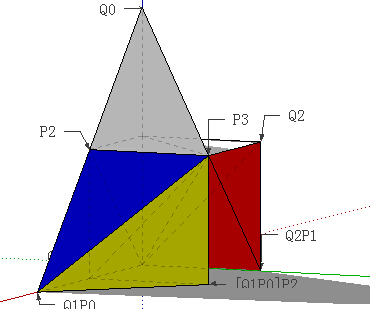}
\includegraphics[width=0.25\textwidth]{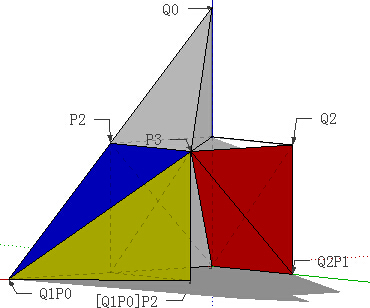}
\includegraphics[width=0.25\textwidth]{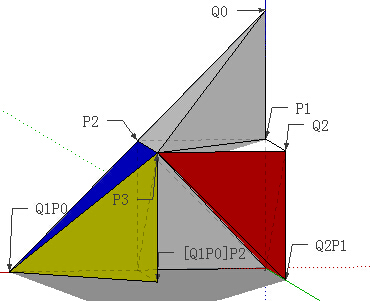}
\hspace*{\fill}\vspace{-0.2cm}
\caption{\small{Seven Simplex Tetrahedrons assembled (termed as Class $7_{2}$ \#\textbf{A}), viewing from different angles. The labels in each vertex help us to distinguish the tetrahedrons. The colored tetrahedrons means the three added tetrahedrons on the Kernel Simplex. }}\label{Fig:class72A}
\end{figure}

\begin{figure}[!htp]
\includegraphics[width=0.25\textwidth]{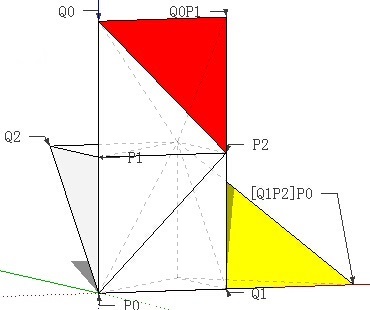}
\includegraphics[width=0.25\textwidth]{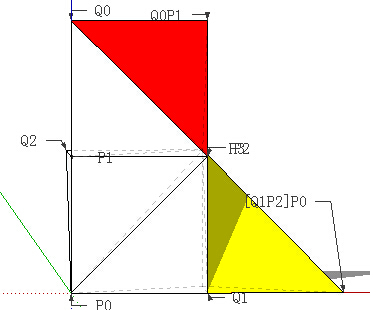}
\includegraphics[width=0.25\textwidth]{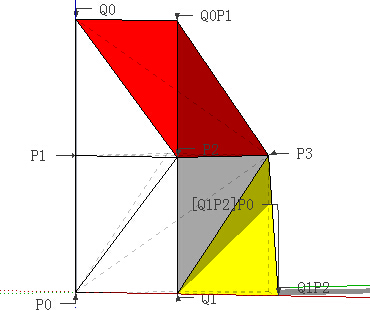}
\includegraphics[width=0.25\textwidth]{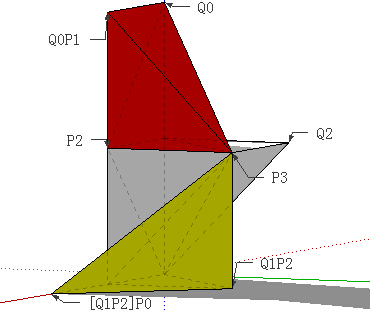}\\
\includegraphics[width=0.25\textwidth]{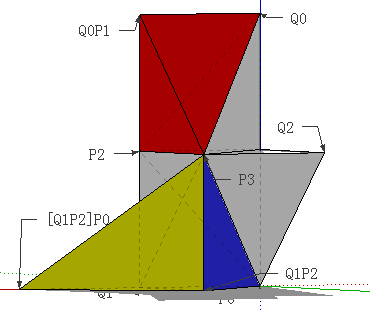}
\includegraphics[width=0.25\textwidth]{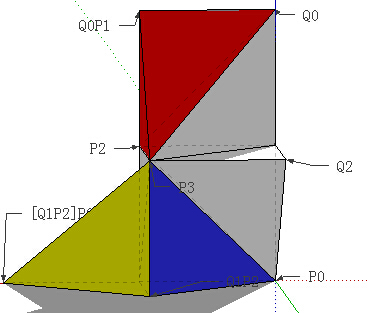}
\includegraphics[width=0.25\textwidth]{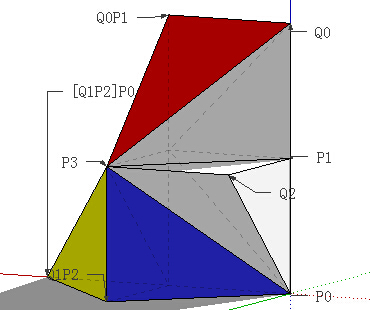}
\includegraphics[width=0.25\textwidth]{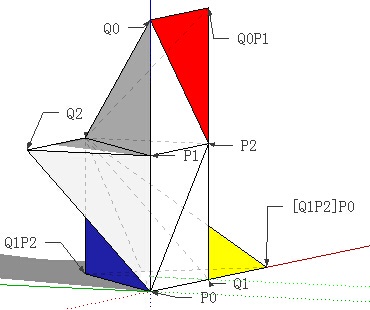}
\hspace*{\fill}\vspace{-0.2cm}
\caption{\small{Seven Simplex Tetrahedrons assembled (termed as Class $7_{2}$ \#\textbf{B}), viewing from different angles. The labels in each vertex help us to distinguish the tetrahedrons. The colored tetrahedrons means the three added tetrahedrons on the Kernel Simplex. }}\label{Fig:class72B}
\end{figure}

The following step is to classify two non-isometric models generated from $7_{2}$. You can see the 3D view in Figure \ref{Fig:class72A} and \ref{Fig:class72B}.
\noindent\begin{itemize}
\item Class $7_{2}$ \#\textbf{A}, three mirror points:
$Q_2P_1=\{0, 1, 0\},~Q_{1}P_{0}=\{2, 0, 0\},~[Q_{1}P_{0}]P_{2}=\{1, 1, 0\}.$
\small{\begin{itemize}
\item Simplex 4: $\{P_0, Q_2P_1, Q_2, P_3\}$,\qquad $0\le x \le z\le y \le 1  $;
  \item Simplex 5: $\{Q_1P_{0},Q_{1},P_{2},P_{3}\}$,\qquad $0\le y \le z\le 2-x \le 1 $;
  \item Simplex 6: $\{Q_1P_{0},Q_{1},[Q_1P_{0}]P_{2},P_{3}\}$,\qquad $0\le z \le y\le 2-x \le 1  $.
 \end{itemize}}\normalsize
\item Class $7_{2}$ \#\textbf{B}, three mirror points:
$Q_0P_1=\{1, 0, 2\},~Q_{1}P_{2}=\{1, 1, 0\},~[Q_{1}P_{2}]P_0=\{2, 0, 0\}$.
\small{\begin{itemize}
\item Simplex 4: $\{Q_0, Q_0P_1, P_2,P_3\}$,\qquad $0\le y \le 2-z\le x \le 1  $;
  \item Simplex 5: $\{P_{0},Q_{1},Q_{1}P_{2},P_{3}\}$,\qquad $0\le z \le y\le x \le 1 $;
  \item Simplex 6: $\{[Q_{1}P_{2}]P_{0},Q_{1},Q_{1}P_{2},P_{3}\}$,\qquad $0\le z\le y\le 2-x \le 1  $.
 \end{itemize}}\normalsize
 \end{itemize}

\begin{figure}[!htp]
\begin{center}
{\includegraphics[width=1\textwidth]{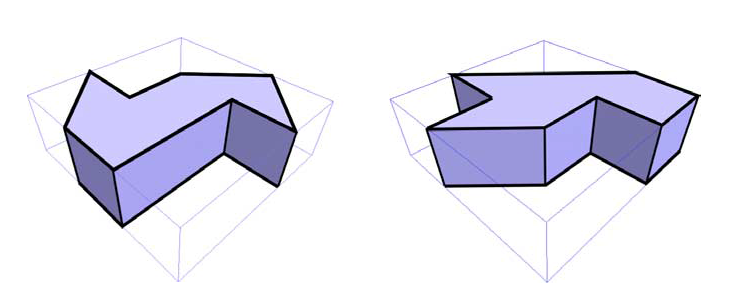}}
\vspace{-0.5cm}
\caption{\small{The GWW prisms once appeared in \cite{ShapeDNA}\cite{Moorhead2012}}.  }\label{Fig:GWWprism}
\end{center}\vspace{-0.2cm}
\end{figure}

 The 3D views in Figure \ref{Fig:class72A} and \ref{Fig:class72B} shows Class $7_{2}$ \#\textbf{A} and $7_{2}$ \#\textbf{B} are non-isometric. Please also see Table \ref{Table:72} for the approximated eigenvalues of this 3D iso-spectral pair. The maximum difference and $L_{2}$-norm error are as follows.
\[\max_{1\leq k \leq 25}|\lambda_{k}-\mu_{k}|=0.6821 \times 10^{-12}, ~~~\|\lambda_{k}-\mu_{k}\|_{L_{2}}=1.4167 \times 10^{-12}.\]
Figure \ref{Fig:GWWprism} shows the GWW prisms which are iso-spectral. They are just constructed from GWW by sweeping the face into a third orthogonal direction for the same distance.
\begin{table}[!htp]
  \small{\begin{center}
  \caption{\small{The first 25 approximated eigenvalues of the 3D iso-spectral pair: Class $7_{2}$ \#\textbf{A} and Class $7_{2}$ \#\textbf{B}, using the simple finite difference method (please also see \cite{Moorhead2012}) with mesh size $h=1/20$. The Difference refers to the absolute difference between the eigenvalues.}}\vspace{-0.3cm}
    \begin{tabular*}{\linewidth}{@{\extracolsep{\fill}}*{4}{c}}                                   \toprule \toprule
   Eigenvalue    &Class $7_{2}$ \#\textbf{A}   &Class $7_{2}$ \#\textbf{B}   &Difference ($\times 10^{-12}$)   \\\midrule
 1  & 44.9835 &  44.9835 &  0.1137\\
 2  & 61.4888 &  61.4888 &  0.1208\\
 3  & 70.0240 &  70.0240 &  0.3837\\
 4  & 80.6794 &  80.6794 &  0.1705\\
 5  & 86.2937 &  86.2937 &  0.2700\\
 6  &102.1243 & 102.1243 &  0.1990\\
 7  &104.7903 & 104.7903 &  0.0711\\
 8  &110.7750 & 110.7750 &  0.1847\\
 9  &121.5084 & 121.5084 &  0.2416\\
 10 &124.1022 & 124.1022 &  0.1279\\
 11 &129.6075 & 129.6075 &  0.6821\\
 12 &136.1989 & 136.1989 &  0.0568\\
 \large{\color{red}13} &137.0019 & 137.0019 &  0.1421\\
 14 &142.1820 & 142.1820 &  0.2842\\
 15 &146.8989 & 146.8989 &  0.0853\\
 16 &157.7060 & 157.7060 &  0.2842\\
 17 &160.9049 & 160.9049 &  0.3695\\
 18 &163.9460 & 163.9460 &  0.1421\\
 19 &165.5862 & 165.5862 &  0.3411\\
 20 &171.1039 & 171.1039 &  0.3695\\
 21 &178.3842 & 178.3842 &  0.3979\\
 22 &183.0443 & 183.0443 &  0.4547\\
 23 &185.0180 & 185.0180 &  0.1137\\
 24 &191.9694 & 191.9694 &  0.2274\\
 25 &195.6234 & 195.6234 &  0.0284\\
  \bottomrule
 \end{tabular*}\label{Table:72}
  \end{center}}
\end{table}

\begin{table}[!htp]
  \small{\begin{center}
  \caption{\small{The first 25 approximated eigenvalues of the 3D iso-spectral pair: Class $7_{3}$ \#\textbf{A} and Class $7_{3}$ \#\textbf{B}, using the simple finite difference method ( please also see \cite{Moorhead2012}) with mesh size $h=1/20$. The Difference refers to the absolute difference between the eigenvalues.}}\vspace{-0.3cm}
    \begin{tabular*}{\linewidth}{@{\extracolsep{\fill}}*{4}{c}}                                   \toprule \toprule
   Eigenvalue    &Class $7_{3}$ \#\textbf{A}   &Class $7_{3}$ \#\textbf{B}   &Difference ($\times 10^{-12}$)   \\\midrule
 1 &49.2289  & 49.2289& 0.0213 \\
 2 & 56.0467 &  56.0467& 0.0284 \\
 3 & 72.6396 &  72.6396& 0.1847 \\
 4 & 79.9743 &  79.9743&0.1137\\
 5 & 92.0586 &  92.0586&0.1847\\
 6 & 99.5111 &  99.5111&0.3837\\
 7 &104.0452 & 104.0452&0.2274\\
 8 & 113.2988 & 113.2988&0.2984\\
 9 &120.5720 & 120.5720& 0.1137\\
 10 &124.4357 & 124.4357&0.5542\\
 11 &131.7220 & 131.7220&0\\
 12 &133.3562 & 133.3562& 0.3411\\
 \large{\color{red}13} &136.1989 & 136.1989&0.3695\\
 14 &144.0266&  144.0266& 0.4547\\
 15 &152.4877&  152.4877& 0.1705\\
  16&156.3340&  156.3340& 0.1137\\
 17 &156.5645 & 156.5645& 0.0568\\
 18 &162.8653 & 162.8653& 0.3979\\
 19& 169.5866 & 169.5866& 0.0853\\
 20 &173.2912 & 173.2912& 0.3695\\
 21& 179.6543 & 179.6543& 0.0853\\
 22& 185.0656 & 185.0656& 0.3979\\
23 & 186.5775 & 186.5775& 0.4263\\
 24& 190.4249&  190.4249&0.1137\\
25 & 193.5350 & 193.5350&0.6821\\
    \bottomrule
 \end{tabular*}\label{Table:73}
  \end{center}}
\end{table}

\begin{figure}[!htb]
\includegraphics[width=0.5\textwidth]{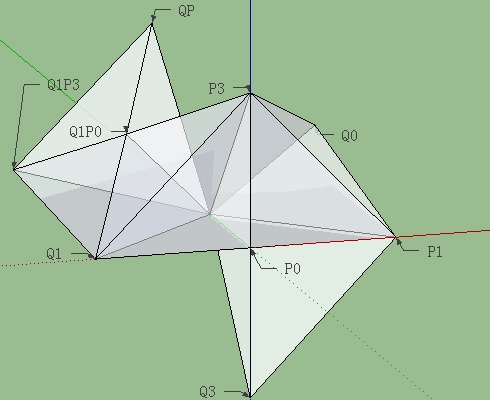}
\includegraphics[width=0.5\textwidth]{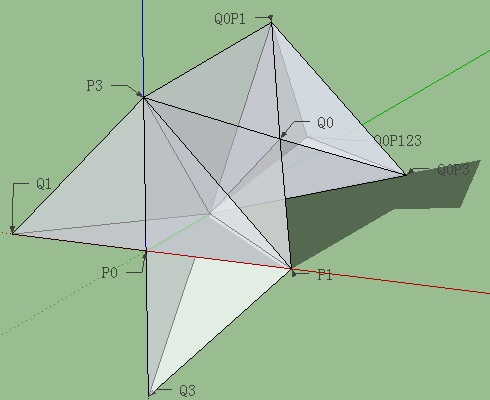}
\hspace*{\fill}\vspace{-0.2cm}
\caption{\small{Two 3D models assembled by seven Wall Tetrahedrons, viewing in the 3D X-ray form.  Generated from $7_{3}$ by fixing the face $(P_{0},P_{1},P_{3})$, this pair is iso-spectral and non-isometric.}}\label{Fig:73wallAB}
\end{figure}

Table \ref{Table:73} shows the detailed results of the first 25 approximated eigenvalues of  Class $7_{3}$ \#\textbf{A} and Class $7_{3}$ \#\textbf{B}. Similarly, we also calculate the maximum difference and $L_{2}$-norm error.
\[\max_{1\leq k \leq 25}|\lambda_{k}-\mu_{k}|= 0.6821 \times 10^{-12}, ~~~\|\lambda_{k}-\mu_{k}\|_{L_{2}}=1.6766 \times 10^{-12}.\]
\begin{figure}[!htb]
\includegraphics[width=0.5\textwidth]{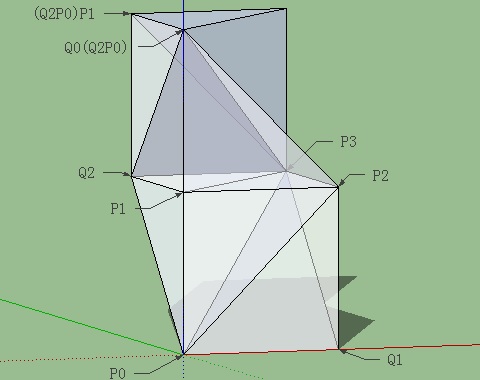}
\includegraphics[width=0.5\textwidth]{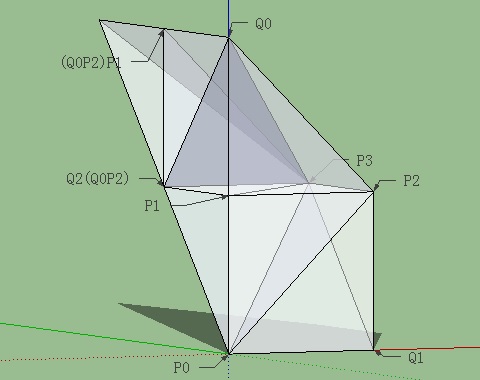}
\hspace*{\fill}\vspace{-0.2cm}
\caption{\small{Seven Simplex Tetrahedrons assembled (termed as Class $7_{3}$ \#\textbf{A} and $7_{3}$ \#\textbf{B}), viewing in 3D X-ray form. Because of symmetry in Simplex Tetrahedron, faces $(P_{1},P_{3},Q_{2})$ and $(P_{1},P_{3},Q_{0}P_{2})$ coincide in Class $7_{3}$ \#\textbf{A}, faces $(P_{1},P_{3},Q_{0})$ and $(P_{1},P_{3},Q_{2}P_{2})$ coincide in Class $7_{3}$ \#\textbf{B}.}}\label{Fig:class73AB}
\end{figure}
 Figure \ref{Fig:class73AB} depicts the 3D X-ray form of Class $7_{3}$ \#\textbf{A} and $7_{3}$ \#\textbf{B}. At the same time, we also illustrate other 3D iso-spectral models constructed by Wall Tetrahedrons. Different from the construction of Class $7_{3}$ \#\textbf{A} and $7_{3}$ \#\textbf{B}, we fix or forget face $(P_{0},P_{1},P_{3})$ and do multilevel mirror reflection following the rule of $7_{3}$.
\noindent Given four 3D vertices of the Basic Wall Tetrahedrons with coordinate:
\[P_0= \{0, 0, 0\},~P_1= \{1, 0, 0\},~P_2 = \{0, 1, 0\},~P_3=\{0, 0, 1\}.\]
We can calculate the mirror points according to Equation (\ref{Equation:mirror1}) and Equation (\ref{Equation:mirror2}). Figure \ref{Fig:73wallAB} depicts the two 3D iso-spectral models.
\begin{remark}\rm
For Basic Simplex Tetrahedron ($0\le y\le x\le z\le 1$) , there exists ``analytical'' eigenmodes (please refer the explanation of the GWW's eigenmodes in \cite{T.ADriscoll1997}). The first eigenmode is expressed in Simplex Tetrahedron of $\mathbf{(1^2+2^2+3^2)\large{\pi}}$. The first approximated eigenmodes calculated by finite difference method are marked in red color, see Table 1-3. Numerical techniques are therefore employed to approximate such eigenvalues, but even this has its own problems (e.g. the way to deal with reentrant corners).
\end{remark}

\begin{remark}\rm
Note that if the colors of faces are permuted the
resulting 3D pairs also become iso-spectral. For example, the pair $7_{1}$ represents three
distinguishable pairs of models for a fixed tetrahedron.
\end{remark}

\section{Conclusion}

\noindent
Milnor's work is not a direct answer to the original version
of Kac's question because it is not concerned with the planar domains. But from
Milnor's example one learned that there really exists non-congruent domains with the same eigenvalue spectrum. From a mathematical point of view, the spectrum of a drum corresponds to the set of eigenvalues of the negative Laplacian on a given planar domain, where the solutions vanish at the border (Dirichlet boundary conditions). Sunada's idea was to reduce the problem of finding iso-spectral manifolds to a group-theoretical problem, namely,
constructing triplets of groups having a certain property. Based on the Sunada's method, Buser et al. constructed 17 families of iso-spectral pairs, and confirmed their iso-spectrality by the transplantation method. Giraud and Thas reviewed the highlighting mathematical and physical aspects of iso-spectrality.

We revisit the serials of reflection rule inherent in the 17 families of planar iso-spectral domains. Following the reflection rule visualized by red-blue-black, many 3D iso-spectral models can be constructed. Especially for the tetrahedron as the basic building block, we present some visualized 3D pairs which are iso-spectral and non-isometric. This extension of constructing iso-spectral pairs can also be used in higher dimensions, with transplantation method to guarantee the respective iso-spectrality.

%
%
%

\section{Acknowledgements}
This work was supported by
the National Science Foundation of China (NSFC 91230109). The first author would like to thank Professor Olivier Giraud for illuminating email
indication and would also like to thank Professor Hua Li, introducing me the software Google SketchUp to draw the 3D Figures.
\appendix
\section{Appendix }

 First, we show the details of how to obtain the transplantation matrix of Class $7_{1}$ \#\textbf{A} and \#\textbf{B}.
\begin{itemize}
\item Class $7_{1}$ \#\textbf{A}, seven Simplex Tetrahedrons labelled by $i,~i=0,1,\ldots,6$.
\small{\begin{itemize}
\item Face: $\{P_1, P_2, P_3\}=\{P_1, Q_2, P_3\}$,\qquad (0~1)~~~(3~6);
  \item Face: $\{P_0, P_2, P_3\}=\{Q_0, P_2, P_3\}$,\qquad (0~2)~~~(1~4);
  \item Face: $\{P_0, P_1, P_3\}=\{P_0, Q_1, P_3\}$,\qquad (0~3)~~~(2~5).
 \end{itemize}}\normalsize
\item Class $7_{1}$ \#\textbf{B}, seven Simplex Tetrahedrons labelled by $i,~i=0,1,\ldots,6$.
\small{\begin{itemize}
\item Face: $\{P_1, P_2, P_3\}=\{Q_1, P_2, P_3\}$,\qquad (0~1)~~~(2~5);
  \item Face: $\{P_0, P_2, P_3\}=\{P_0, Q_2, P_3\}$,\qquad (0~2)~~~(3~6);
  \item Face: $\{P_0, P_1, P_3\}=\{Q_0, P_1, P_3\}$,\qquad (0~3)~~~(1~4).
 \end{itemize}}\normalsize
 \end{itemize}
\[\small{A^1=\left(
\begin{array}{ccccccc}
 0 & 1 & 0 & 0 & 0 & 0 & 0 \\
 1 & 0 & 0 & 0 & 0 & 0 & 0 \\
 0 & 0 & 1 & 0 & 0 & 0 & 0 \\
 0 & 0 & 0 & 0 & 0 & 0 & 1 \\
 0 & 0 & 0 & 0 & 1 & 0 & 0 \\
 0 & 0 & 0 & 0 & 0 & 1 & 0 \\
 0 & 0 & 0 & 1 & 0 & 0 & 0 \\
\end{array}
\right),A^2=\left(
\begin{array}{ccccccc}
 0 & 0 & 1 & 0 & 0 & 0 & 0 \\
 0 & 0 & 0 & 0 & 1 & 0 & 0 \\
 1 & 0 & 0 & 0 & 0 & 0 & 0 \\
 0 & 0 & 0 & 1 & 0 & 0 & 0 \\
 0 & 1 & 0 & 0 & 0 & 0 & 0 \\
 0 & 0 & 0 & 0 & 0 & 1 & 0 \\
 0 & 0 & 0 & 0 & 0 & 0 & 1 \\
\end{array}
\right),A^3=\left(
\begin{array}{ccccccc}
 0 & 0 & 0 & 1 & 0 & 0 & 0 \\
 0 & 1 & 0 & 0 & 0 & 0 & 0 \\
 0 & 0 & 0 & 0 & 0 & 1 & 0 \\
 1 & 0 & 0 & 0 & 0 & 0 & 0 \\
 0 & 0 & 0 & 0 & 1 & 0 & 0 \\
 0 & 0 & 1 & 0 & 0 & 0 & 0 \\
 0 & 0 & 0 & 0 & 0 & 0 & 1 \\
\end{array}
\right)};\]
\[\small{B^1=\left(
\begin{array}{ccccccc}
 0 & 1 & 0 & 0 & 0 & 0 & 0 \\
 1 & 0 & 0 & 0 & 0 & 0 & 0 \\
 0 & 0 & 0 & 0 & 0 & 1 & 0 \\
 0 & 0 & 0 & 1 & 0 & 0 & 0 \\
 0 & 0 & 0 & 0 & 1 & 0 & 0 \\
 0 & 0 & 1 & 0 & 0 & 0 & 0 \\
 0 & 0 & 0 & 0 & 0 & 0 & 1 \\
\end{array}
\right),B^2=\left(
\begin{array}{ccccccc}
 0 & 0 & 1 & 0 & 0 & 0 & 0 \\
 0 & 1 & 0 & 0 & 0 & 0 & 0 \\
 1 & 0 & 0 & 0 & 0 & 0 & 0 \\
 0 & 0 & 0 & 0 & 0 & 0 & 1 \\
 0 & 0 & 0 & 0 & 1 & 0 & 0 \\
 0 & 0 & 0 & 0 & 0 & 1 & 0 \\
 0 & 0 & 0 & 1 & 0 & 0 & 0 \\
\end{array}
\right),B^3=\left(
\begin{array}{ccccccc}
 0 & 0 & 0 & 1 & 0 & 0 & 0 \\
 0 & 0 & 0 & 0 & 1 & 0 & 0 \\
 0 & 0 & 1 & 0 & 0 & 0 & 0 \\
 1 & 0 & 0 & 0 & 0 & 0 & 0 \\
 0 & 1 & 0 & 0 & 0 & 0 & 0 \\
 0 & 0 & 0 & 0 & 0 & 1 & 0 \\
 0 & 0 & 0 & 0 & 0 & 0 & 1 \\
\end{array}
\right)}.\]
\normalsize
By the definition \ref{Def:} in Section \ref{Sec:Transplantation method}, we obtain all the commutation relations (permutation matrices $A^\nu$ and $B^\nu$). Finding $T$ just amounts to solving a system of Equation (\ref{Equation:TABT}).

\begin{figure}[!htp]
\begin{center}
{\includegraphics[width=1\textwidth]{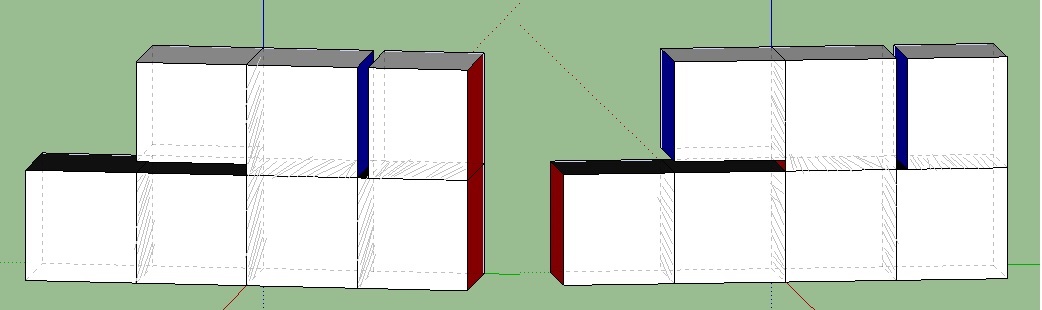}}
\vspace{-0.5cm}
\caption{\small{Seven unit cubes assembled (generated from  Class $7_{1}$ \#\textbf{A}), viewing in two angles. The colors on faces help us to perform mirror reflection operation.}  }\label{Fig:cube2}
\end{center}\vspace{-0.2cm}
\end{figure}

\begin{figure}[!htp]
\begin{center}
{\includegraphics[width=1\textwidth]{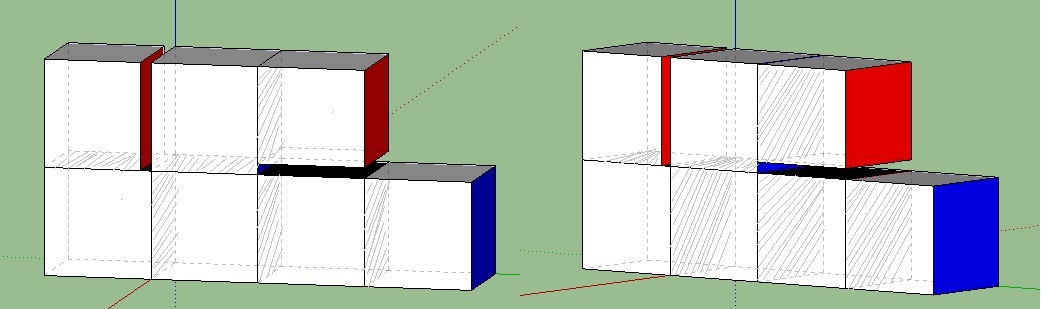}}
\vspace{-0.5cm}
\caption{\small{Seven unit cubes assembled (generated from  Class $7_{1}$ \#\textbf{B}), viewing in two angles.  The colors on faces help us to perform mirror reflection operation.}  }\label{Fig:cube3}
\end{center}\vspace{-0.2cm}
\end{figure}

Second,we present two isometric models generated from the Class $7_{1}$. They are constructed by seven unit cubes depicted in Figure \ref{Fig:cube2} and \ref{Fig:cube3}. The gaps between some cubes are to be interpreted as having zero boundary.
\begin{figure}[!htp]
\begin{center}
{\includegraphics[width=1\textwidth]{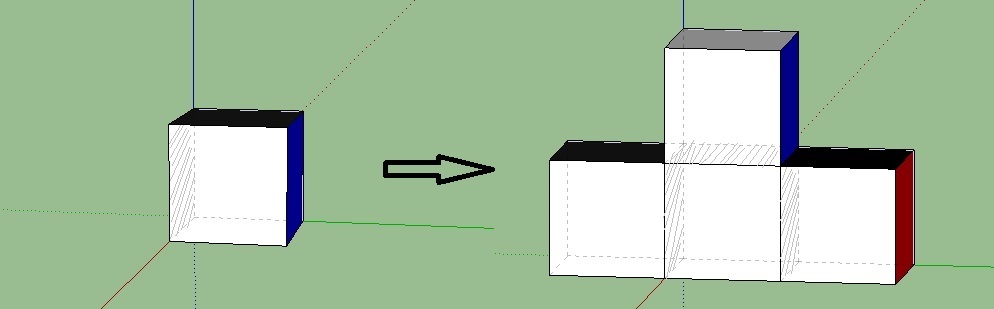}}
\vspace{-0.5cm}
\caption{\small{One unit cube and four unit cubes assembled (termed as Kernel Unit Cube). The colors on faces help us to perform mirror reflection operation.}  }\label{Fig:cube1}
\end{center}\vspace{-0.2cm}
\end{figure}

\noindent

\end{document}